\documentclass[12pt,draftcls,peerreviewca,onecolumn,a4paper,dvips]{IEEEtran}
\usepackage{pifont}
\usepackage{amsfonts}
\usepackage{mathrsfs}
\usepackage{amsfonts}
\usepackage{amssymb}
\usepackage{graphicx}
\usepackage{psfrag}
\usepackage{amsmath}
\usepackage{array}
\usepackage{cases}
\usepackage{color}
\begin{document}

\author{\IEEEauthorblockN{Liangzhong Ruan,  stevenr@ust.hk \\ Vincent K.N. Lau,
eeknlau@ee.ust.hk}
\IEEEauthorblockA{Dept. of Electrical and Electronic Engineering\\
the Hong Kong University of Science and Technology\\
Clear Water Bay, Kowloon, Hong Kong} }
\title{Decentralized Dynamic Hop Selection and Power Control in Cognitive Multi-hop  Relay Systems}

\newtheorem{Thm}{Theorem}[section]
\newtheorem{Lem}{Lemma}[section]
\newtheorem{Asm}{Assumption}[section]
\newtheorem{Def}{Definition}[section]
\newtheorem{Remark}{Remark}
\newtheorem{Prob}{Problem}
\newtheorem{Alg}{Algorithm}
\newtheorem{Cor}{Corollary}[section]
\definecolor{deep-green}{rgb}{0,0,0}
\definecolor{deep-blue}{rgb}{0,0,0}

\maketitle

\begin{abstract}
In this paper,  we consider a cognitive multi-hop relay secondary
user (SU) system sharing the spectrum with some primary users (PU).
The transmit power  as well as the hop selection of the cognitive
relays can be  dynamically adapted according to the local (and
causal) knowledge of the instantaneous channel state information
(CSI) in the multi-hop SU system. We shall determine a low
complexity, decentralized algorithm to maximize the average
end-to-end throughput of the SU system with dynamic spatial reuse.
The problem is challenging due to the decentralized requirement as
well as the causality constraint on the knowledge of CSI.
Furthermore, the problem belongs to the class of stochastic {\em
Network Utility Maximization} (NUM) problems which is quite
challenging {\color{deep-blue}\cite{NUM}}. We exploit the time-scale
difference between the PU activity and the CSI fluctuations and
decompose the problem into a master problem and subproblems. We
derive an asymptotically optimal low complexity solution using
divide-and-conquer and {\color{deep-blue} illustrate} that
significant performance gain can be obtained through dynamic hop
selection and power control. The worst case complexity and memory
requirement of the proposed algorithm is ${\cal O}(M^2)$ and ${\cal
O}(M^3)$ respectively, where $M$ is the number of SUs.

\end{abstract}
\section{Introduction}
Cooperative Communication and Dynamic Spectrum Access (DSA) are two
important technologies that drive the evolution of the next
generation wireless systems. For instance, cooperative communication
\cite{broadcast,cooperative commnumcation} exploits the broadcast
nature of the wireless channel and enhances the reliability of the
packet against channel fading and hence, increases the coverage of
wireless systems. There are a lot of works studying multi-hop relay
network. In \cite{dual hop perf1}, the authors analyzed the
performance of a dual-hop relaying communications over fading
channels. Performance bounds of multi-hop relay system is analyzed
in \cite{multi hop bound}. However, these works did not consider
dynamic resource adaptation in the relay system. In \cite{ARF2}, the
authors investigated the minimum energy per bit treating both
capacity and power consumption as optimization parameters in the
wireless ad-hoc network. The minimization of the transmit power
under the assumption of orthogonal transmissions was studied in
\cite{ARF,ARF4}, in which the optimal parallel-relay channel power
allocation for {\color{deep-blue} \emph{Amplify and Forward} (AF) and \emph{Decode and Forward} (DF)} were derived.  However, in all these works,
the power control solution adapts on the path loss only and failed
to exploit the dynamic fluctuations of microscopic fading. In
\cite{Kim:2008}, the authors considered dynamic power control for
multi-hop relay but the solution is centralized and requires
knowledge of the global channel state information about the entire
adhoc network, which is very difficult to realize in practice.
Furthermore, a fixed number of hops to deliver a packet to the
destination is always assumed in the above works.  Due to the {\em
store-and-forward} penalty in the end-to-end throughput of multi-hop
relaying, it is not always optimal to involve a fixed number of hops
in the multi-hop network. To tackle this issue, various {\em
opportunistic} multi-hop relaying protocols were proposed in
\cite{CU1,CU2,CU3}.  In these designs, the number of hops to deliver
a packet to the destination node changes dynamically according to
the channel conditions. However, in these works, the opportunistic
multi-hop protocols are heuristic in nature and the performance is
studied by simulation and empirical measurements. In \cite{one relay
perf}, performance analysis on one-hop relay protocol is {\color{deep-blue}studied.} In
\cite{CU4,CU5}, performance analysis on some simple opportunistic
multi-hop relaying protocols {\color{deep-blue}is studied}. Furthermore, they all
assume constant transmit power and deterministic channels where the
effects of random fading is ignored.

On the other hand, DSA is an important new paradigm of spectrum
access in which a secondary system dynamically shares {\color{deep-blue}medium} with a
higher priority primary system. Using {\em cognitive radios} {\color{deep-blue}(CRs)}
\cite{CR1,CR2}, the nodes in the secondary user (SU) systems  sense
the activity of the primary users (PUs) and access the spectrum only
if the primary system is idle. In other words, the SU system
dynamically share the spectrum with the PU systems by exploiting the
burstiness of the PU traffic in the temporal, frequency and spatial
domains. One key issue of DSA or {\color{deep-blue}CR} is the efficiency
of spectrum sharing between the SU and PU systems.  In
\cite{CR3,CR4}, the authors considered a {\color{deep-blue}CR} system
based on the {\em interference avoidance approach} in which the SU
could transmit only if there are no active PUs within the coverage
of the SU system. While such approach exploits the burstiness of the
PU activity without requiring the knowledge of PU signal structure,
the access opportunity of the SU system will be quite low for SU
separated by a large distance as such access opportunity exists only
if all the PU along the SU coverage are idle simultaneously.  As a
result, cognitive multi-hop relay for the SU systems is a promising
solution to resolve this issue of low probability of access for
distant secondary users. While intuitively, cognitive multi-hop relay
could significantly enhance the spectrum sharing efficiency between
the SU and PU systems, there are still a number of technical
challenges to overcome as listed below.

\begin{itemize}
\item {\bf Jointly Optimal Opportunistic Hop Selection and Power Control for Cognitive Multi-hop relays:}
Most of the existing works only considered either the power control
\cite{ARF,ARF4} or the opportunistic multi-hop relaying protocols.
It is very important to jointly optimize both the forward hopping
strategy and the power control policy to exploit the instantaneous
fluctuations of PU activities and the microscopic fading in order to
improve the performance of the cognitive multi-hop relays.

\item {\bf Dynamic Spatial Reuse in Cognitive Multihop Relaying:}
In most of the existing works studying power control or forward
hopping in multihop relay \cite{CU1,CU2,CU3}, they focus entirely on
the multihop aspects of the problem and assume that the multi-hop
network does not have to share spectrum with any PU systems. This
simplifies the problem significantly. While
this is a reasonable assumption in the regular multihop network
without PU, such symmetric spatial reuse is not always possible in
cognitive multihop relay network due to the random PU activities on
any hops.

\item {\bf Decentralized Solution with Local Knowledge of Channel State Information (CSI):}
An additional level of difficulty in solving the forward hopping and
power control problem is the requirement of decentralized solution.
In practice, it is very difficult to obtain and keep track of an
up-to-date knowledge of the instantaneous channel state information
for the entire multi-hop network. As a result, it is desirable to
have a decentralized solution which requires knowledge of local
(rather than global) channel state information only. In
\cite{distributedchain}, the authors considered a distributed
resource management scheme for multi-hop CR networks
but no power control is considered and the solution is based on
heuristic design.

\item {\bf Causal Knowledge of Channel States in the Multi-hop Relay Network:}
In most of the existing works \cite{Kim:2008}, not only global
knowledge but also non-causal\footnote{Causality here refers to
whether the source knows about the future channel states along the
entire multihop transmission event from the source to the
destination. In existing works, one way to justify the "non-causal
knowledge" is to assume the channel state remains quasi-static
across the sum of frame durations in the multihop transmission from
the source to the destination.} knowledge of channel states in the
multi-hop network is assumed. Specifically, at $t=0$, the
centralized controller is assumed to have knowledge of all the
channel states in all the hops of the entire multi-hop relay
network. However, by the time the packets are delivered in the
$n$-th hop, the actual channel state may have changed and the
constraint of having causal knowledge of channel states have not
been taken into account in the previous works of power optimization
in multihop relay network.
\end{itemize}

In this paper, we shall try to address the above technical
challenges. We consider a cognitive multi-hop SU system with a
source, a destination and $M$ half-duplex cognitive relays scattered
between the source and the destination. The SU system dynamically
shares the spectrum with a PU system (with many PU nodes). The
transmit power of the SU nodes as well as the hopping sequence of
the cognitive relays are adaptive according to the local (and
causal) knowledge of channel states in the multi-hop SU system to
optimize the average end-to-end throughput. The solution also
accommodates dynamic spatial reuse across the cognitive multi-hop
system. The problem belongs to the class of stochastic
NUM\footnote{Stochastic NUM refers to a Network Utility Maximization
problem where the objective function involves expectation w.r.t. the
stochastic system state and the optimization variables involve not
just actions at a given system state realization but rather a
collection of actions for all system state realizations. This is a
challenging problem because of the huge dimension of variables
involved as well as the lack of explicit closed form expression for
the objective function in terms of the control policy.} problems,
which is well-known to be challenging. To obtain a decentralized
solution for the throughput optimization problem we exploit the
time-scale difference between the PU activity and the CSI
fluctuations and decompose the problem into a master problem and
subproblems (operating at different time scales). To deal with the
causality requirement\footnote{In our paper, we allow the CSI to be
time varying across different hops in the multi-hop transmission and
the control policy is adaptive to the current information (but not
the future CSI knowledge) only.}, we express the subproblems into
recursive forms  and solve them using {\em divide-and-conquer}. We
show that significant performance gains on the throughput of the SU
system can be obtained using joint forward hopping and power control
over a wide range of PU activity. Furthermore, we show that the
decentralized solution has worst case complexity of ${\cal O}(M^2)$
and is asymptotically optimal for large $M$.

\section{System Model, Control Policy and End-to-End Throughput}
\label{sect2} 
\begin{figure}[ht]
\centering
\includegraphics[width=3.8in, height = 2.2 in ]{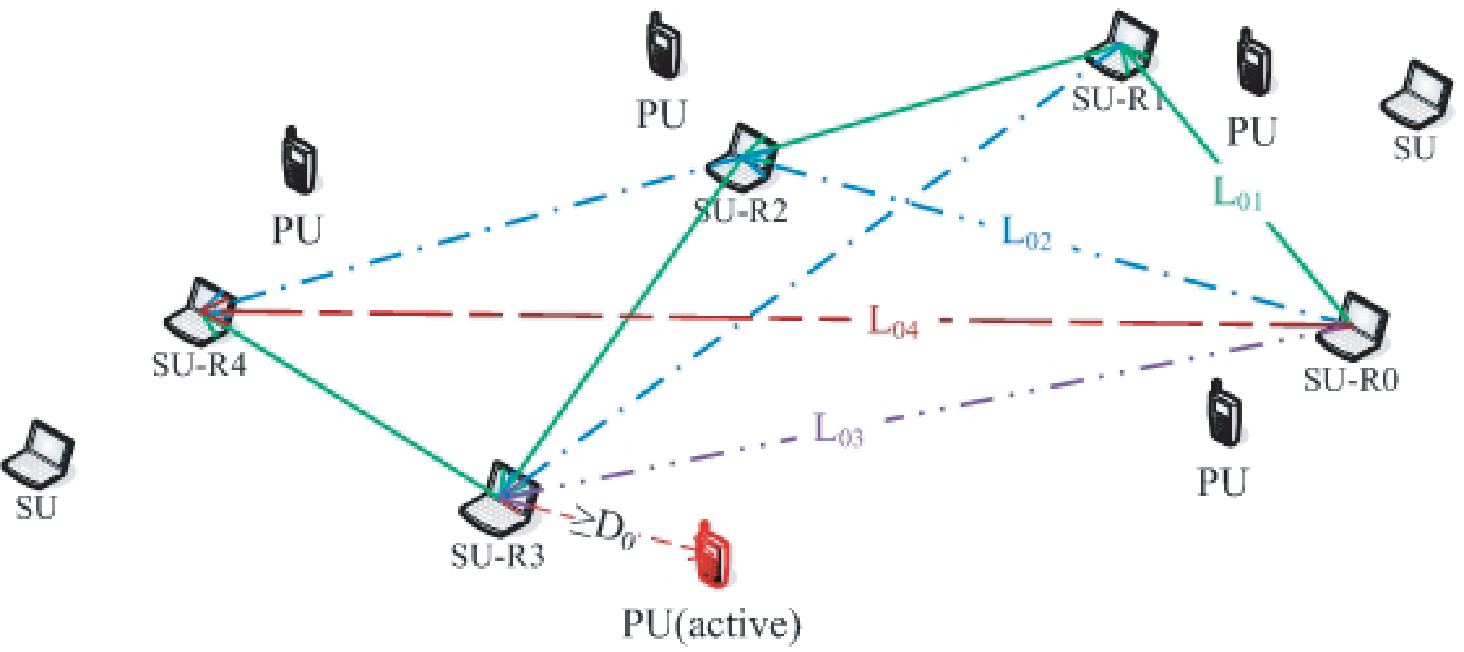}
\caption{System Architecture of the Cognitive Multi-hop Relay
Network. PU and SU denote the Primary User and the Secondary User,
respectively. The source node in the SU system delivers packet to
the destination node via the help of the linear multi-hop relays.
Each node has a cognitive radio to detect and sense the local PU
activity. The nodes are numbered according to the transmission route
determined by certain Layer 3 protocol. }\label{fig:system}
\end{figure}

Fig.\ref{fig:system} illustrates the system model of
the cognitive multi-hop relay system.  The SU system consists of a
cognitive source, a destination and several randomly distributed
relays.
\begin{Asm} The system adopts certain Layer 3 ({\color{deep-blue}network layer}) protocol
to determine a route from the SU source node to the SU destination
node, where the {\color{deep-blue}route} is defined as a sequence of ordered
nodes $\mathbb{R}=<R_0,R_1,...R_M>$, where $R_0, R_M$ are source
node and destination node respectively. This route is assumed to be
fixed throughout the communication session.
\end{Asm}

  Denote
the source as $R_0$, destination $R_M$ and $M-1$ cognitive relays,
$\{R_1,...,R_{M-1}\}$, which are distributed between $R_0$ and
$R_M$. The PU system consists of short-range wireless systems where
the PU nodes are assumed to distribute uniformly (with a density of
$\rho_p$) over the SU coverage area. Each of the PU node is assumed
to have bursty activity with an active probability of $P_a$. The PU
and the SU systems share common frequency spectrum and the SU system
{\color{deep-blue}can} access the channel only when all the involved PU nodes are
idle. In the following, we shall elaborate
on the channel model, control policy and the end-to-end throughput
of the SU cognitive multi-hop relay system.

\subsection{Channel Model}
\begin{figure}[ht]
\centering
\includegraphics[width=4.5in, height = 3.6 in ]{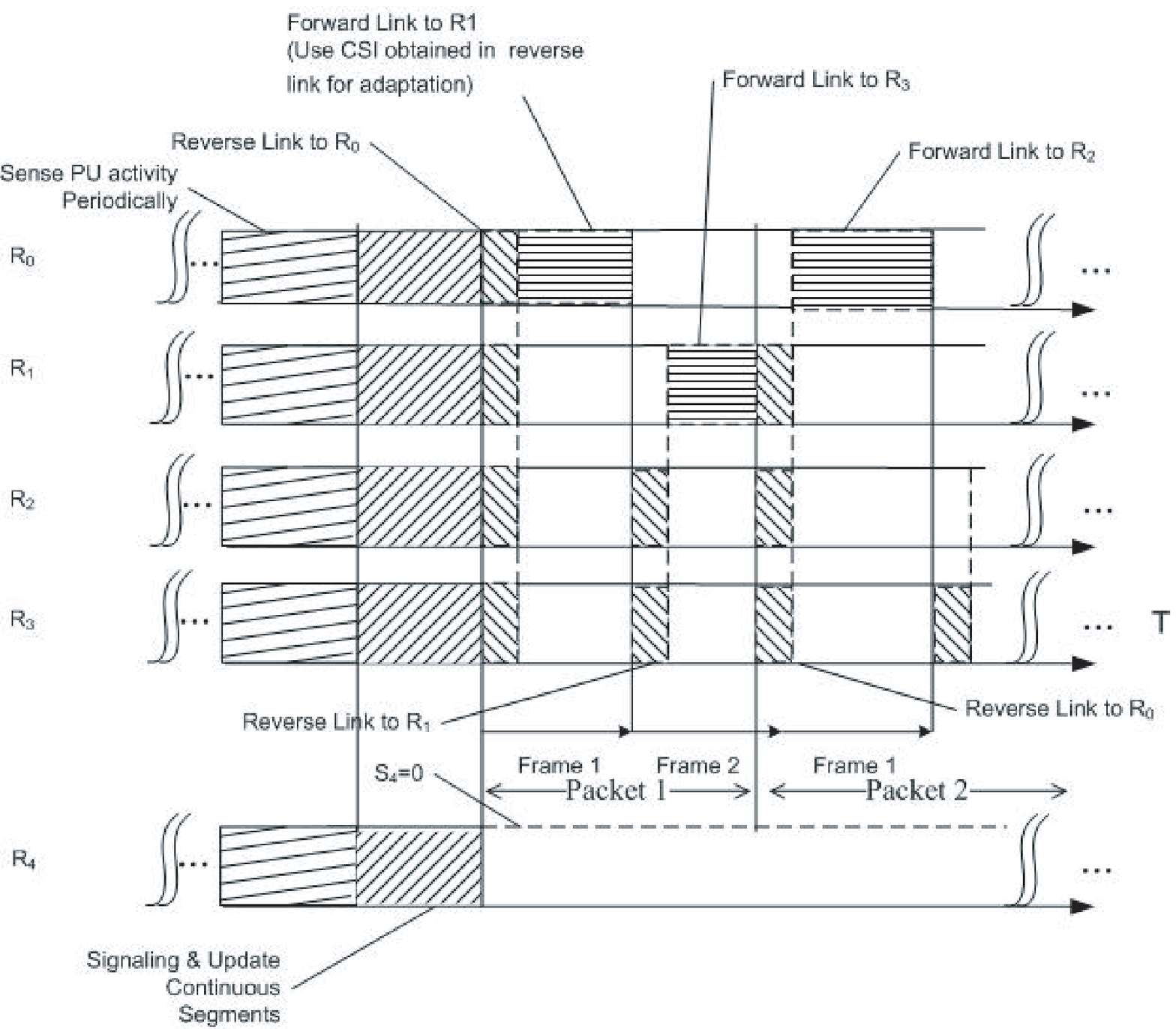}
\caption{Signaling flow of the Cognitive Multi-hop Relay Network. PU
activity is obtained in the periodic sensing frame. The transmitting
node obtains instantaneous local channel state from the reverse
link. Although each hop may have a different frame duration, such
design can be accommodated over a synchronous relay network. For
example, similar to IEEE 802.16j, each relay node in the system is
synchronized to the symbol boundary. As a result, the time varying
frame duration (quantized to the integral number of symbols) can be
realized on top of the symbol-synchronized relay network.
}\label{fig:obtainCSI}
\end{figure}

Figure~\ref{fig:obtainCSI} illustrates the signaling flow in
multi-hop relay system. For the sensing of PU activity, we adopt the
distributed sensing and centralized data fusion model as in IEEE
802.22. For instance, there are periodic quiet periods in the SU
system that enable the sensing of PU activity. During the quiet
periods, the {\color{deep-blue}SUs} sense the PU activity locally and sends the
sensing results to {\color{deep-blue}the other} SU nodes. The SUs exchange the sensing
results and update \emph{continuous segment} (to be defined in the
next subsection) information for data fusion. Define $A_m\in\{0,1\}$
as the sensing result which represents the availability of the
shared spectrum to the SU system ($A_m=1$ denotes that the shared
spectrum is available to SU node $R_m$) and
$\mathbf{A}=(A_1,...,A_m)$ be the vector of PU activity states for
the $M$ SU nodes. We assume an SU node $R_m$, {\color{deep-blue}$m=\{0,1...,M\}$} has
access ($A_m=1$) to the shared spectrum {\color{deep-blue}if and only if} the nearest active PU
node is at least $D_0$ meters away from the SU node\footnote{$D_0$
is determined by the mean interference constraint to PU. For
instance, denote $P_{int}$ as the interference constraint from SU to
PU, $P_0$ is the mean transmitting power of SU, then
$D_{0}\ge\left(\frac{P_0}{P_{int}}\right)^{\frac{1}{\alpha}}$, where
$\alpha$ is the path loss factor.}. Furthermore, assume that $\mathbf{A}$ remains quasi-static between
{\color{deep-blue}two consecutive} sensing periods. This is a reasonable assumption as the
burstiness of the PU nodes are of a longer time scale compared to
the packet frame duration.

The received signal at the $j$-th SU node from the $i$-th SU node at
the $k$-th frame is given by:
\begin{eqnarray}
Y_{ij}(k)={\color{deep-blue} H_{ij}(k)\sqrt{D_{ij}}} X_{ij}(k) + Z_{ij}
\end{eqnarray}
{\color{deep-blue}where} $X_{ij}(k)$ is the transmitted data symbol from node $i$ to
node $j$, $Z_{ij}$ is the zero-mean complex Gaussian channel noise
(with normalized variance 1) and $G_{ij}(k)=|H_{ij}(k)|^2 D_{ij}$ is
the combined channel loss (including both the large-scale path loss
$D_{ij}$ and the microscopic fading $H_{ij}$)  between node $i$ and
$j$. The microscopic fading $H_{ij}$ is modeled as zero-mean,
unit-variance complex Gaussian i.i.d (independent for different
users) random variables. Let $\mathbf{G}(k)=\{G_{ij}(k):i\neq j,
i,j\in\{0,1,...,M\}\}$ be the global channel state (GCS) information.
We assume $\mathbf{G}$ is quasi-static within a frame. For practical considerations, we have the
following restrictions on the knowledge of the channel states.
\begin{itemize}
\item {\bf Local Knowledge of Channel States:} We assume each of the SU node only has knowledge of the local channel state (LCS, to be defined below) and global PU activity state $\mathbf{A}$
 (which remains quasi-static between
two consecutive sensing periods).
\item {\bf Causal Knowledge of Channel States:}
We assume that each SU node only has causal knowledge of the LCS
and cannot predict into the future.
\end{itemize}

Specifically, we assume at the $k$-th frame, SU node m only has
knowledge about the current LCS:
 $\mathbf{G}_m(k)$. \textcolor{deep-green}{Here, $\mathbf{G}_m(k)=\{G_{mi}(k),i\in\{m+1,...,j\}\}$ in which $j$
 should satisfy:
 $ S_{m+1} = \ldots = S_{j} =1,
S_{j+1} = 0$
  is the local CSI at the $k$-th frame.}
\subsection{System State, Hopping and Power Control Policy, System State Transition Kernel.}
In this section, we shall formally define the control policy in the
cognitive multi-hop relaying system. The multi-hop relay network
operates in a DF manner with half-duplex constraint. At each frame,
the upstream SU node transmits a packet of $B$ bits to its
down-stream nodes using a transmit power which could be dynamically
adjusted based on the current LCS knowledge. The down-stream SU
node(s) attempt to decode the $B$-bits packet before it {\color{deep-blue}can} forward
to the next hop.

In this paper we consider dynamic spatial
reuse in the cognitive multi-hop relay system  as illustrated in Figure~\ref{fig:relay
scheme}. For any given PU states
$\mathbf{A}$, the multi-hop relay chain will be partitioned into
several segments, which is defined as:
\begin{Def}[Continuous Segment in route $\mathbb{R}$]
\label{def:segment} A continuous segment $L_{ij}$ in the cognitive
multi-hop relay chain is defined as a sequence of nodes
$<R_i,...,R_j> \subseteq \mathbb{R}$ such that:
\begin{eqnarray}S_{i-1} =0, S_{i} = \ldots = S_{j} =1, S_{j+1} = 0,
\;\; i,j\in\{1,2...,M\} \label{eqn:conseg}
\end{eqnarray}

(Define $S_{-1} = S_{M+1} = 0$). \ The nodes $R_i$ and $R_j$ are
called the {\em head-node} and the {\em end-node} of the continuous
segment respectively. Define the probability of $\{R_i,...,R_j\}$
forms a continuous segment as $\Pr(i,j)=\Pr(S_{i-1} =0, S_{i} =
\ldots = S_{j} =1, S_{j+1} = 0)$.
~\hfill \IEEEQED \end{Def}

 Spatial reuse is allowed only for relays in different
segments of the partition. Hence, relays in different segments can
transmit different information simultaneously without interfering
each other. Packets are stored at the end-node of each continuous
segment and the end-node are not allowed to transmit except when the
down-stream PU activity becomes idle. However, for relays in one
continuous segment, they have to obey the TDMA constraint and cannot
transmit different information simultaneously at any given time.

{\color{deep-blue} Within a continuous segment $L_{ij}$  induced by
the PU activity $\mathbf{A}$, we shall define the hopping and power
control policy
 as follows:

\begin{Def}[System State of Segment $L_{ij}$]
Suppose $R_i\sim R_j$ induced by a continuous segment $L_{ij}$ under
a PU activity state $\mathbf A$. System state of $L_{ij}$ at frame
index\footnote{\color{deep-blue} The frame index $k$ is equal to the
number of hops already experienced by the packet currently
transmitting in a continuous segment and will be reset to 1 when
this packet is successfully delivered to the end node. Hence, $k$
might be different from segment to segment.} $k\in\{1,2...j-i\}$ is
given by:
 $\eta_{ij}(k)=\{s_{ij}(k),\mathbf{G}_{s_{ij}(k)}\}$, where
$s_{ij}(k)\in\{i,i+1,...j\}$  denotes the index of the source node
at frame $k$, $s_{ij}(1)=i$; $\mathbf{G}_{s_{ij}(k)}$ is the LCS at
node $s_{ij}(k)$. ~\hfill \IEEEQED \end{Def}

\begin{Def}[Control Policy $\Omega_{ij}$ in Segment $L_{ij}$]
A stationary policy  $\Omega_{ij}$ is a mapping from the current system state $\eta_{ij}(k)$ to the corresponding hopping and power control actions.
 The policy $\Omega_{ij} = \{\mathcal{L}_{ij}, \mathcal{P}_{ij}\} $, where:
\begin{itemize}
\item{\em Forward hopping policy $\mathcal{L}_{ij} $}: $l_{ij}(k)=\mathcal{L}_{ij}(\eta_{ij}(k)), k\in \{1,2...j-i\} $,
where the hopping control action (destination node index at frame $k$)   has to satisfy the constraint: $s_{ij}(k)\le l_{ij}(k)\le j $, with the left inequality strictly holds when
$s_{ij}(k)<j$.
\item{\em Dynamic power control policy $\mathcal{P}_{ij} $}: $P_{ij}(k)=\mathcal{P}_{ij}(\eta_{ij}(k)),k\in \{1,2...j-i\} $, where the power control action
(transmitting power at frame $k$) shall satisfy $P_{ij}(k)>0$.
\end{itemize}
~\hfill \IEEEQED \end{Def}

\begin{Def}[System State Transition Kernel]
The source node of at the $k+1$-th frame  $s_{ij}(k+1)$, is determined by the hopping control action in the previous frame $l_{ij}(k)$. Furthermore,
the distribution of the channel state $\mathbf{G}_{s_{ij}(k+1)}$ is independent of the previous system states $\eta_{ij}(k)$
due to the casual knowledge assumption. Hence, the {\em state transition kernel} of the system state $\{\eta_{ij}(k)\}$ is given by:
\begin{eqnarray}
\Pr(\eta_{ij}(k+1)|\eta_{ij}(k),\Omega_{ij})=\mathbf{1}\left(s_{ij}(k+1)=l_{ij}(k)\right)\cdot\Pr(\mathbf{G}_{s_{ij}(k+1)})\label{eqn:kernel}
\end{eqnarray}
~\hfill \IEEEQED \end{Def}
}

\begin{Remark}
Strictly speaking, the forward hopping policy $\mathcal L$ does not
contain all possible hopping sequences w.r.t. a given route
$\mathbb{R}$. For example, potential loops (e.g. $R_i\rightarrow
R_j\rightarrow R_i$) are excluded. Note that it is an intractable
problem to optimize w.r.t. general hopping policies (including
loops) due to the enormous possible policies involved. Instead, we
shall restrict to forward hopping policy only and from which, we
could exploit the structure in the policy space to derive much
simpler solutions.
~\hfill \IEEEQED\end{Remark}

\subsection{End-to-End Throughput with Dynamic Spatial Reuse and Forward Hopping Control}
In order for a SU node to forward a packet, in
any {\em continuous segment}, the node itself must be able to decode
the packet first (DF). Suppose a node is able to decode if and only
if the total mutual information received is no less than B bits.
Hence, we have:
\begin{eqnarray}
T_{ij}(k)\cdot\log(1+ G_{s_{ij}(k)l_{ij}(k)}(k)
P_{ij}(k)) \geq B, \;\; k \in
\{1,2...j-i\}, s_{ij}<j\label{eqn:info}
\end{eqnarray}

where $i,j$ satisfy \eqref{eqn:conseg} and $T_{ij}(k)$ is the
transmitting time of the $k$-th frame in \emph{continuous segment}
$L_{ij}$. {\color{deep-blue} We first formally define the {\em
per-hop reward and cost} below.
\begin{Def}[Per Hop Reward and Per Hop Cost] Define the {\em reward} at the $k$-th frame as  the time
taken to transmit 1 bit at the $k$-th frame:
\begin{eqnarray}T(\eta_{ij}(k),\Omega_{ij})=\left\{\begin{array}{l}\frac{1}{\log(1+
G_{s_{ij}(k)l_{ij}(k)}(k) P_{ij}(k))}\mbox{ when: } s_{ij}(k)<j
\\ 0 \mbox{ otherwise.}
\end{array}\right. \end{eqnarray}
Define the cost at the $k$-th frame as the power consumed to transmit 1 bit at the $k$-th frame:
\begin{eqnarray}P(\eta_{ij}(k),\Omega_{ij})=\left\{\begin{array}{l}\frac{P_{ij}(k) }{\log(1+
G_{s_{ij}(k)l_{ij}(k)}(k) P_{ij}(k))}\mbox{ when:
}s_{ij}(k)<j\\0\mbox{ otherwise.}\end{array}\right. \end{eqnarray}
~\hfill \IEEEQED \end{Def}

Note that $T_{ij}(k)=B\cdot T(\eta_{ij}(k),\Omega_{ij})$ and hence the average data
rate in the {\em continuous segment} $L_{ij}$ can be expressed as:
\begin{eqnarray}
U_{ij}= E^{\Omega_{ij}}\left(\frac{B}{\sum_{k\in\{1,2...j-i\},s_{ij}(k)<j}T_{ij\_k}}\right)
=\label{eqn:av-thp-sub1}E^{\Omega_{ij}}
\left(\frac{1}{\sum_{k=1}^{j-i}T(\eta_{ij}(k),\Omega_{ij})}\right)
\end{eqnarray}

where the expectation $E^{\Omega_{ij}}$ is taken w.r.t. the probability measure {\em induced} by the control policy $\Omega_{ij}$ and the transition kernel in \eqref{eqn:kernel}.
Similarly, average power consumption $\overline{P}_{ij}$ in $L_{ij}$
can be expressed as:
\begin{equation}
\overline{P}_{ij}=E^{\Omega_{ij}}\left(\frac{\sum_{k=1}^{j-1}P(\eta_{ij}(k),\Omega_{ij})}{\sum_{k=1}^{j-i}T(\eta_{ij}(k),\Omega_{ij})}\right)\label{eqn:av-subpower}
\end{equation}
}

The end-to-end average throughput of the cognitive multi-hop system
can be written as the weighted sum of average data rate of all {\em
continuous segments} with end-node $R_M$:
\begin{equation}
\label{eqn:av-thp1}\overline{U}(\Omega) =
\sum_{i=0}^{M-1} \Pr(i,M) U_{iM}
\end{equation}

The average sum-power constraint is given by:
\begin{equation}
\label{eqn:tx-pwr1} \sum_{i=0}^{M-1}\sum_{j=i+1}^{M}\Pr(i,j)
\overline{P}_{ij}\leq P_0
\end{equation}

Moreover, the conventional flow-balance constraint\footnote{The conventional flow balance constraint ensures that the output flow does
not exceed the input flow at any SU node.} is given by:
\begin{eqnarray}
\label{eqn:flow-balance1} \sum_{i=0}^{m-1}\Pr(i,m) U_{im}&\ge&
\sum_{j=m+1}^{M}\Pr(m,j) U_{mj} \;\forall m \in \{1 ,..., M-1\}
\end{eqnarray}

\section{Problem Formulation}
\begin{figure}[ht]
\centering
\includegraphics[width=3.4in, height = 1.7in ]{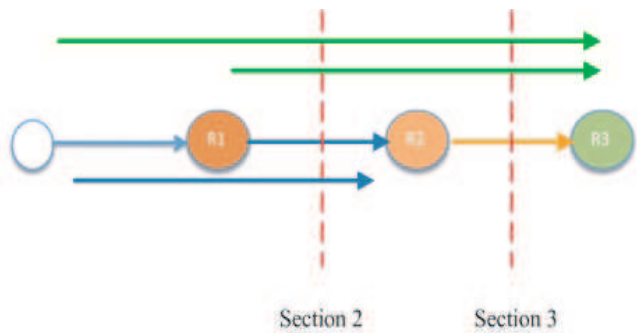}
\caption{Illustration of the {\em Section Flow Balance} criteria.}\label{fig:sectionbalance}
\end{figure}

{\color{deep-blue}
Note that the conventional flow-balance constraint in \eqref{eqn:flow-balance1} may not be convex\footnote{\color{deep-blue} A convex (concave) function subtracting another convex (concave) function
is neither convex nor concave in general.}.
} To solve this issue, we introduce {\color{deep-blue}a}
new balance criteria, namely the {\em section flow-balance
criteria}. For instance, we consider the sum of average data rate
passing through each section (rather than each node). Specifically,
the sum-average data rate passing through the $m$-th section ($m \in
\{1,2,...M\}$ as illustrated in Fig.\ref{fig:sectionbalance}).
Define: $ \bar{U}_m = \sum_{i=0}^{m-1}\sum_{j=m}^{M}\Pr(i,j)
U_{ij}$. The {\em section flow-balance criteria} is given by:
\begin{equation}
\label{eqn:flow-balance2} \overline{U}_m \ge \overline{U}_{m+1},\;
\forall m \in \{1 ,..., M-1\}
\end{equation}

In the following lemma, we shall illustrate that the section
flow-balance criteria is in fact equivalent to the conventional
per-node flow-balance:
\begin{Lem} \label{Lem:flow-balance} [{\em Equivalence of the flow balance criteria}] The conventional {\em per-node} flow balance constraint in
(\ref{eqn:flow-balance1}) is equivalent to the {\em per-section}
flow balance criteria in (\ref{eqn:flow-balance2}).
\end{Lem}
\proof please refer to Appendix \ref{PLem:flow-balance} for the
proof.
\endproof

Lemma \ref{Lem:flow-balance} gives an equivalent form for
traditional flow-balance criteria. Moreover, note that the objective
$\overline{U}((\Omega))$ in (\ref{eqn:av-thp1}) is equal
to:
\begin{eqnarray}
\nonumber \overline{U}(\Omega) &=& \sum_{i=0}^{M-1}
\Pr(i,M) U_{iM} \;= \sum_{i=0}^{M-1} \sum_{j=M}^{M} \Pr(i,j)
U_{ij}\;= \bar{U}_{M}
\\ \label{eqn:av-thp2} &=& \min(\{\bar{U}_{1}, \bar{U}_{2},...,\bar{U}_{M}
\}) \mbox{ (Due to section flow balance criteria
(\ref{eqn:flow-balance2}))}
\end{eqnarray}
where $\Omega$ is the overall control policy: $\Omega=\{\Omega_{ij}, \forall i,j$ that satisfies \eqref{eqn:conseg} under a PU actively state $\mathbf{A}\}$.

From (\ref{eqn:av-thp2}), the optimization problem can be formulated
as:
\begin{Prob}[Original Problem]\label{prob:org}
\begin{eqnarray}
&&\label{eqn:obj_2}\bar{U}=\max_{\Omega}\left[\min_{m\in\{1,..., M\}}
\sum_{i=0}^{m-1}\sum_{j=m}^{M}\Pr(i,j){U}_{ij}\right]
\\&&\nonumber  \mbox{Subject to:}
\\\label{eqn:power_2} && \sum_{i=0}^{M-1}\sum_{j=i+1}^{M}\Pr(i,j)
\overline{P}_{ij}\leq P_0
\end{eqnarray}

where: ${U}_{ij}$, $\overline{P}_{ij}$ is given by
\eqref{eqn:av-thp-sub1} and \eqref{eqn:av-subpower} respectively.
\end{Prob}


\subsection{Decomposition of Main Problem}
The optimization problem in ~\eqref{eqn:obj_2} is too complex to
solve directly. Furthermore, due to the causality constraint in the
control policies ${\cal P}$ and ${\cal L}$, the solution is not
trivial and brute-force solution will not lead to viable solutions.
However, it is worthy noting that for a given PU activity state
$\mathbf{A}$, operations on different \emph{continuous segment} are
naturally separated from each other. (e.g. as in Fig~\ref{fig:relay
scheme}, when $S_4=0$, hopping and power control policy in segment
$R_0\sim R_3$ has no direct {\color{deep-blue} influence} on that in $R_5\sim R_6$).
Making use of this insight, we shall first decompose the problem
into a {\em master problem} and a {\em sub-problem}. Define:
${\cal{P}}_{main}=\{\overline{P}_{ij}\}, \;i,j \in \{0,1,...,M\},
\;i<j$.

We have the following decomposition theory:
\begin{Lem}
Optimization problem consisting of a master problem (
Problem~\ref{prob:mas}, with ${\cal{P}}_{main}$ as the optimization
policy) and $\frac{M(M-1)}{2}$ subproblems (Problem~\ref{prob:sub},
with ${\cal L}_{ij},{\cal P}_{ij}$ as the optimization policies) is
equivalent to Problem~\ref{prob:org}.

\begin{Prob}[Master Problem]\label{prob:mas}
\begin{eqnarray}
&&\label{eqn:obj_3}\overline{U}=\max_{{\cal
P}_{main}}\left[\min_{m\in\{1,..., M\}}
\sum_{i=0}^{m-1}\sum_{j=m}^{M}\Pr(i,j){U}^*_{ij}(\overline{P}_{ij})\right]
\\&&\nonumber  \mbox{Subject to:}
\\\label{eqn:power_3b} && \sum_{i=0}^{M-1}\sum_{j=i+1}^{M}\Pr(i,j) \overline{P}_{ij}\leq P_0
\end{eqnarray}
\end{Prob}

\begin{Prob}[Subproblem]\label{prob:sub}
\begin{eqnarray}
\label{eqn:obj_sub_1} &&{U}^*_{ij}(\overline{P}_{ij}) =\max_{\mathcal{L}_{ij},\mathcal{P}_{ij}}E^{\Omega_{ij}}
\left(\frac{1}{\sum_{k=1}^{j-i}T(\eta_{ij}(k),\Omega_{ij})}\right)
\\&&\nonumber  \mbox{Subject to:}
\\\label{eqn:power_sub_1} &&E^{\Omega_{ij}}\left(\frac{\sum_{k=1}^{j-1}P(\eta_{ij}(k),\Omega_{ij})}{\sum_{k=1}^{j-i}T(\eta_{ij}(k),\Omega_{ij})}\right)\le\overline{P}_{ij}
\end{eqnarray}
\end{Prob}
\label{thm:decompose}
\end{Lem}
\proof Please refer to Appendix~\ref{Pthm:decompose} for the proof.
\endproof
\section{Decentralized Hop Selection and Power Control Algorithm}
\subsection{Solving the Sub Problem}
To satisfy the causality constraint of the  control policy on the
local CSI, we have to model the subproblem in a recursive form so as
to apply dynamic programming (DP) \cite{DPbook}. However, problem
(\ref{eqn:obj_sub_1}) cannot be expressed in a recursive form and
hence, could not be divide-and-conquered. To tackle the challenges,
we shall solve a lower bound version of the problem. We shall show
that the lower bound solution is indeed asymptotically tight for
large number of nodes. {\color{deep-blue}
\subsubsection{\textbf{Asymptotically Optimal Solution}}
We first elaborate a suboptimal solution for the subproblem (Problem~\ref{prob:sub}).
 Let \begin{eqnarray}
\Omega_{ij}^{LB}= \arg\min_{\Omega_{ij}}E^{\Omega_{ij}} \left[\sum_{k=1}^{j-1}
T(\eta_{ij}(k),\Omega_{ij}) +
\lambda_{ij}\left(P(\eta_{ij}(k),\Omega_{ij})-\overline{P}_{ij}T(\eta_{ij}(k),\Omega_{ij})\right)
\right] \label{eqn:LB3}
\end{eqnarray}  where the parameter $\lambda_{ij}$ in the suboptimal solution $\Omega_{ij}^{LB}$ is given by the roots of the equation\footnote{For any given $\lambda_{ij}$, $\Omega^{LB}_{ij}$ is determined by \eqref{eqn:LB3}. Substitute both
policy to the \eqref{eqn:power_sub_3}, the LHS become a function of
$\lambda_{ij}$}:
\begin{equation}
\label{eqn:power_sub_3}
E^{\Omega_{ij}}\left(\frac{\sum_{k=1}^{K_{ij}}P(\eta_{ij}(k),\Omega_{ij})
}{\sum_{k=1}^{K_{ij}}T(\eta_{ij}(k),\Omega_{ij})}\right)
=\overline{P}_{ij}
\end{equation}

Note that the solution $\Omega_{ij}^{LB}$ is a feasible but
suboptimal solution of the subproblem (Problem~\ref{prob:sub}). We
have the following lemma about the property of the suboptimal
solution $\Omega_{ij}^{LB}$.

\begin{Lem}[Asymptotic Optimality of $\Omega^{LB}_{ij}$]
\label{Lem:lagrange}
If the following conditions are satisfied: 1)
For any $\epsilon>0$, there exists a finite $C>0$ such that when
$|s-t|\geq C$, $G_{st} < \epsilon$; and 2) $G_{st}\ge G_{st'}$,
$G_{st} \ge G_{s't}$ when $t'\ge t>s\ge s'$; then we have: $U_{ij}^{LB}(\overline{P}_{ij}) \rightarrow
U^*_{ij}(\overline{P}_{ij})$, as $|j-i|\rightarrow \infty$.  where $U_{ij}^{LB}(\overline{P}_{ij})$
is the average throughput of the segment $R_i\sim R_j$ using the suboptimal control $\Omega_{ij}^{LB}$. ~\hfill \IEEEQED
\end{Lem}
\proof Please refer to Appendix \ref{pLem:langrange} for the proof.
\endproof
}
\begin{Remark}[{\color{deep-green} Physical} Interpretations of Conditions (1) and (2) in Lemma~\ref{Lem:lagrange}]\label{lab:sub}
The condition 1) in Lemma~\ref{Lem:lagrange} means that the nodes
are not \emph{"over concentrated"} on one spot. This is is a mild requirement,
which only excludes the special topologies where there are infinite number of nodes over a finite coverage area.
 The condition
2) refers to the path loss dominated situations, which applies for medium-range (over 2-5 km) multi-hop networks.
~\hfill \IEEEQED\end{Remark}

{\color{deep-blue}As a result, the suboptimal solution
$\Omega^{LB}_{ij}$ has reasonable performance in general cases (as
will be illustrated in Section V) and it is asymptotically optimal
for large number of nodes. }In order to derive $\Omega^{LB}_{ij}$,
we shall first express
 into a recursive form and solve the problem by
divide-and-conquer using DP. Define
\begin{equation}
\label{eqn:g_1}g(\eta_{ij}(k);P_{ij}(k),l_{ij}(k))=\frac{1+\lambda_{ij}(P_{ij}(k)-\overline{P}_{ij})}{\log(1+P_{ij}(k)G_{s_{ij}(k)l_{ij}(k)}(k))}
\end{equation}
then the problem \eqref{eqn:LB3} can be expressed
recursively as:
\begin{equation}
\label{eqn:obj_sub_DP1}
J(s_{ij}(k))=E_{\mathbf{G}_{s_{ij}(k)}}
[\min_{P_{ij}(k),l_{ij}(k)}(g(\eta_{ij}(k);P_{ij}(k),l_{ij}(k))+J(l_{ij}(k)))]
\end{equation}
where $J(m)$ is called the {\em expected cost} from node $R_m$ to
$R_j$. Note that $J(j)=0$ and $J(s_{ij}(1))=J(i)$ gives the
value of (\ref{eqn:LB3}). As a result of the recursive form in \eqref{eqn:obj_sub_DP1}, the
{\em backward recursion algorithm} to solve problem ~\eqref{eqn:LB3}
is summarized in the following.

\fbox{\parbox{15.6cm}{
\begin{Alg}[Offline and Online Solution of the Sub Problem]$\;$
\label{Alg:sol_sub}
\begin{itemize}
\item {\bf Offline Recursion}:
\begin{itemize}
\item {\bf Step 1:} Initialize $\lambda_{ij} = 0$.
\item {\bf Step 2:} For $s=j-1,j-2,...,i$, determine $J(s)$ by {\color{deep-blue}(Here we assume node $R_s$ has the knowledge of the
distribution of the local channel state $\mathbf{G}_{s}$)}:
\begin{equation}
J(s)= E_{\mathbf{G}_{s}(k)} \min_{m\in\{s+1,...j\}}
\left[\frac{1+\lambda_{ij}(P_s^*(\lambda_{ij})-\overline{P}_{ij})}{\log(1+P_s^*(\lambda_{ij})G_{s,m}(k))}+J(m)\right]
\end{equation} where $P^*_s(\lambda_{ij})$ is the solution to \eqref{eqn:opt_p_sub} defined below. The values of $J(s)$ is stored.
\item {\bf Step 3:} Substitute solution obtained  from Step 2 into (\ref{eqn:power_sub_3}). If the LHS is larger (smaller) than $\overline{P}_{ij}$ by $\epsilon$, increase (decrease) $\lambda_{ij}$ by a step $\delta$ and go to Step 2. Otherwise, stop.
\end{itemize}
\item {\bf Online Policy:}
\begin{itemize}
\item {\bf Step 1:} Set $k=1$ and $s_k=i$.
\item {\bf Step 2:} Obtain the local CSI $\mathbf{G}_{s_{ij}(k)}$ and the optimizing hop selection
and power control actions are given by $P^*_{s_{ij}(k)}(\lambda_{ij})$
and:
\[l_{k}^*=\arg\min_{s\in\{l_k+1,...,j\}}
 \left[\frac{1+\lambda_{ij}(P_{l_k}^*(\lambda_{ij})-\overline{P}_{ij})}
 {\log(1+P_{ij}(k)^*(\lambda_{ij})G_{l_k,s}(k))}+J(s)\right] \]
\item {\bf Step 3:} Set $k:=k+1$, $s_{k+1}=l_{k}^*$. If $s_{k+1}\neq j$, goto Step 2. Otherwise, stop.
\end{itemize}
\end{itemize}
\end{Alg}
}}
\begin{equation}
\label{eqn:opt_p_sub}\frac{G_{s_{ij}(k)l_{ij}(k)}(k)}{(1+P_{ij}(k)G_{s_{ij}(k)l_{ij}(k)}(k))
\log(1+P_{ij}(k)G_{s_{ij}(k)l_{ij}(k)}(k))+(\overline{P}_{ij}-P_{ij}(k))G_{s_{ij}(k)l_{ij}(k)}(k)}
= \lambda_{ij}
\end{equation}

\begin{Remark}
\label{Remark:sub} Note that the memory size of the table in the
offline recursion is $j-i$. The computational complexity for the
online algorithm in each step $k$ is only of the order $j-i$. Hence,
the online algorithm has worst case complexity ${\cal O}(M^2)$ and
worst  case memory requirement ${\cal O}(M)$ for each continuous
segment $i,j$.
~\hfill \IEEEQED\end{Remark}

\subsection{Solving the Main Problem}
After solving for the subproblem, we shall focus on solving the main
problem based on $U^{LB}_{ij}(\overline{P}_{ij})$ (which is of a
longer time scale) in this section. We first establish the following
Theorem regarding the concavity of $U^{LB}_{ij}(\overline{P}_{ij})$
w.r.t. $\overline{P}_{ij}$.

\begin{Lem}[Concavity of the Lower Bounds of $U^*_{ij}(\overline{P}_{ij})$]
\label{Lem:subconcave} The lower bound
($U^{LB}_{ij}(\overline{P}_{ij})$) of $U^*_{ij}(\overline{P}_{ij})$
is a concave function of $\overline{P}_{ij}$.
\end{Lem}
\proof Please refer to Appendix \ref{pLem:subconcave} for the proof.
\endproof

\begin{figure}[ht]
\centering
\includegraphics[scale= 0.38 ]{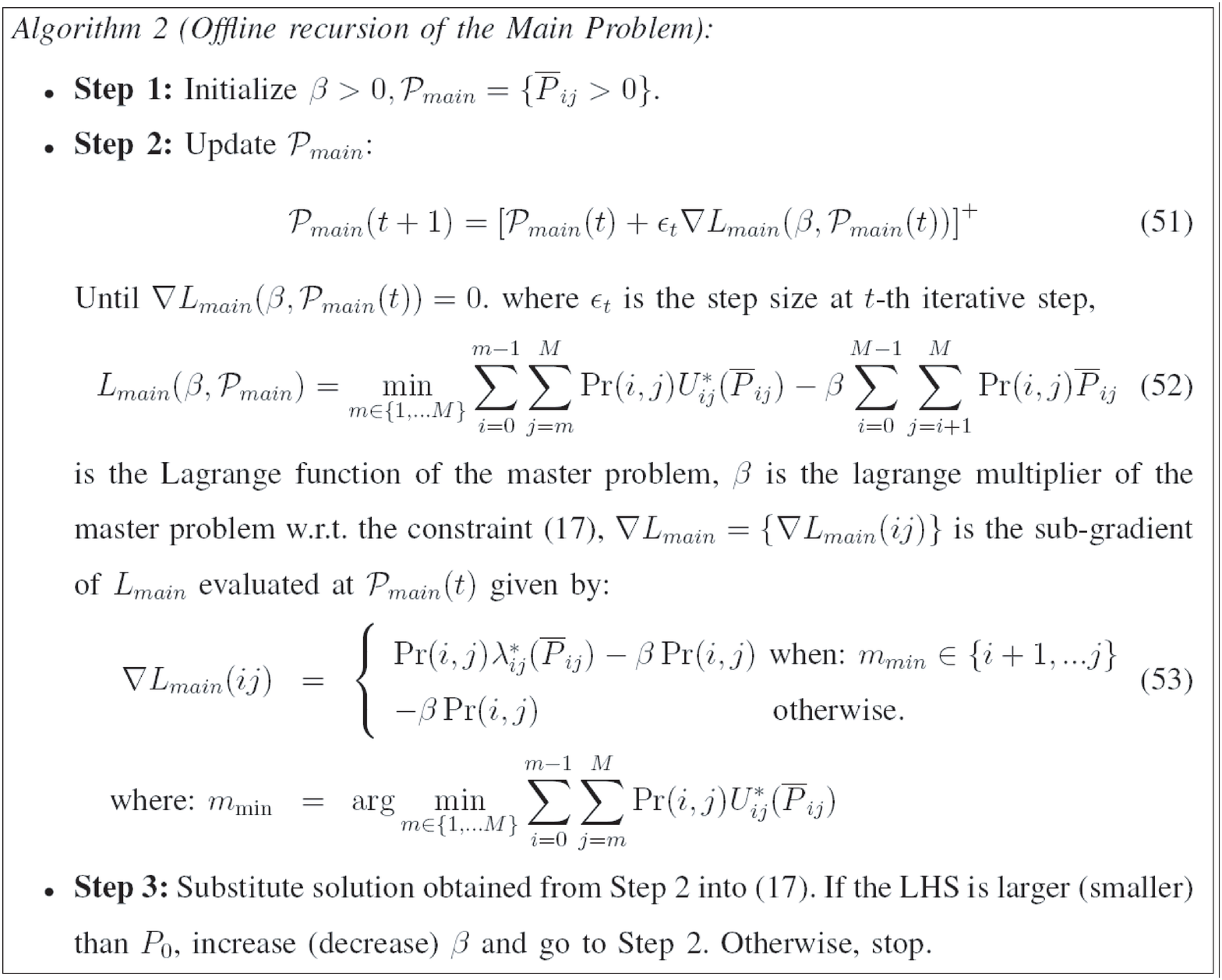}
\caption{ Algorithm description for the Main Problem }\label{fig:alg}
\end{figure}

From Lemma \ref{Lem:subconcave}, it is easy to deduce that the
lower-bound version of the master problem in (\ref{eqn:obj_3}) [with
$U_{ij}^*(\overline{P}_{ij}$) replaced by
$U_{ij}^{LB}(\overline{P}_{ij})$] is a convex optimization problem.
As a result,  the standard gradient search could be applied to solve
the master problem. Please refer to Figure~\ref{fig:alg} for the detailed algorithm description.

\begin{Remark}
Note that the offline recursion needs to be updated only when there
are changes {\color{deep-green} in} the PU statistics or the SU path
loss and in practice, the above offline algorithm is computed  over
a long time scale. Combining the master problem and the subproblems
the total memory requirement of the offline table in algorithm 1 is
$\mathcal {O}(M^3)$. ~\hfill \IEEEQED\end{Remark}

\section{Simulation Results}
In this section, we shall illustrate the performance of the proposed
scheme by simulation. We consider a multi-hop cognitive relay system
with $6$ nodes ($\{R_0,R_1,...R_5\}$) and 6 PUs (one PU in the
neighborhood of each SU node). The distance between $R_0$ and $R_5$
is 5, and the other 4 nodes randomly scatter between them. Path loss
between two nodes $R_i, R_j$ is given by the "flat-earth model"
\cite{flatearth}: $\log_{10}D_{ij}= -\alpha\log_{10}d_{ij} \;(dB)$
where $d_{ij}$ is the distance between the two nodes and $\alpha$ is
the path loss exponent. The proposed scheme is compared with four
schemes below:
\begin{itemize}
\item {\bf Direct transmission only (Baseline 1):} $R_0$ transmit directly to
$R_5$ when all PU remain silent ($S_i = 1, \forall i\{0,1,...5\}$).
This is equivalent to the case without relay.
\item {\bf Per-node transmission only (Baseline 2): } if $R_m$ ($\forall m\{0,1,...4\}$)
received a packet in previous frames, it transmits this packet to
$R_{m+1}$ when the PU activity permits ($A_m = S_{m+1} =1$). This
corresponds to the traditional DF multi-hop relay scheme.
\item {\bf Direct (per-node) transmission with dynamic spatial reuse (Baseline 3/4)}: These two schemes adopt the same dynamic spatial reuse method as the proposed scheme. Yet, within each {\em continuous segment}, they adopt direct and per-node transmission
respectively.
\end{itemize}

\begin{figure}[ht]
\centering
\includegraphics[width=4.5in, height = 3.7in ]{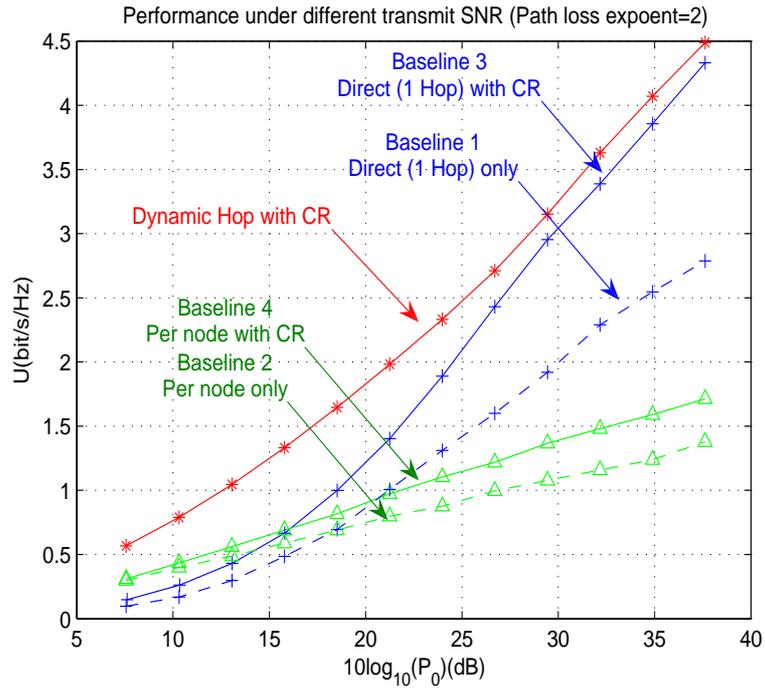}
\caption{Average end-to-end throughput versus transmit SNR $P_0$.
The PU activity is given by $\Pr(A_m=0)=0.15$ and the path loss
exponent is given by $2$. }\label{fig:Main_SNR}
\end{figure}

\begin{figure}[ht]
\centering
\includegraphics[width=4.5in, height = 3.7in ]{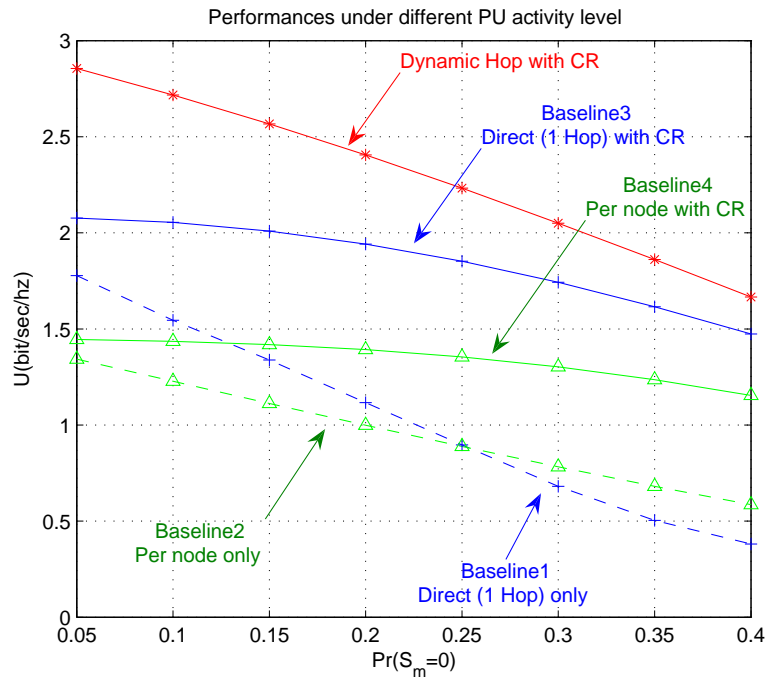}
\caption{Average end-to-end throughput versus PU activity
$\Pr(A_m=0)$. The transmit SNR is 30dB with path loss exponent given
by $3$. }\label{fig:Main_Pr}
\end{figure}

Figure~\ref{fig:Main_SNR} and Figure~\ref{fig:Main_Pr} illustrate
the average end-to-end throughput ($\overline{U}$) versus the
average SNR ($P_0$) and PU activity level ($\Pr(A_m=0)$)
respectively. The proposed scheme achieves significant throughput
gains over a wide range of SNR and PU activities. This gain is
contributed by both the dynamic hop selection as well as dynamic
power control. Comparison with baseline 1 illustrates how cognitive
relay could help to increase the probability of access and
efficiency of spectrum sharing in general. Comparison with baseline
2 and 3 illustrates the importance of joint dynamic power and
opportunistic hop selection in cognitive multihop systems. The gain
contributed by the dynamic hop selection is most significant under
moderate SNR. At very high SNR, the dynamic hopping performance
approaches that of the baseline 3, illustrating the system always
perform one-hop direct transmission to avoid the half-duplex
penalty. At very low SNR, the performance of the proposed scheme
approaches that of the baseline 4, illustrating that the system
prefer hop-by-hop transmission for SNR gain.

\begin{figure}[ht]
\centering
\includegraphics[width=4.5in, height = 3.7in ]{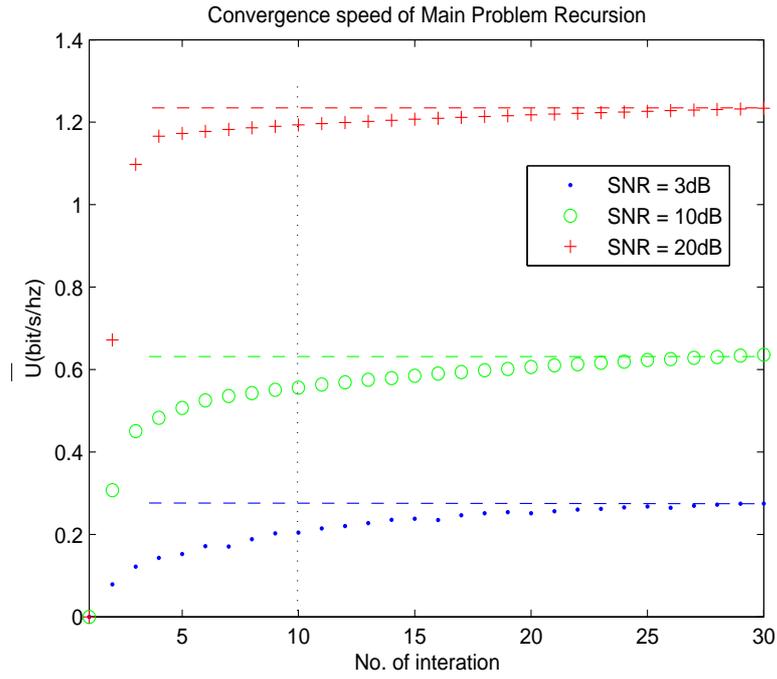}
\caption{Average end-to-end throughput versus Number of iterations
in Algorithm 2. The PU activity is given by $\Pr(A_m=0)=0.15$ and
the path loss exponent is given by $3$. }\label{fig:Main_converge}
\end{figure}

Figure~\ref{fig:Main_Pr} illustrate that the dynamic hopping gain is more prominent under low PU activity.
This is because at low PU activity, there is a higher chance of forming a longer {\em continuous segment} and hence,
 more flexible choices for the dynamic hop selection. Figure~\ref{fig:Main_converge} illustrate the convergence rate of
the off-line recursion for the \emph{Main problem} (Algorithm 2).
The proposed algorithm can achieve $90\%$ of the converged
performance within 10 iterations and converges after about 30
iterations. This iteration efficiency is good enough for off-line
algorithms.

\begin{figure}[ht]
\centering
\includegraphics[width=4.5in, height = 3.7in ]{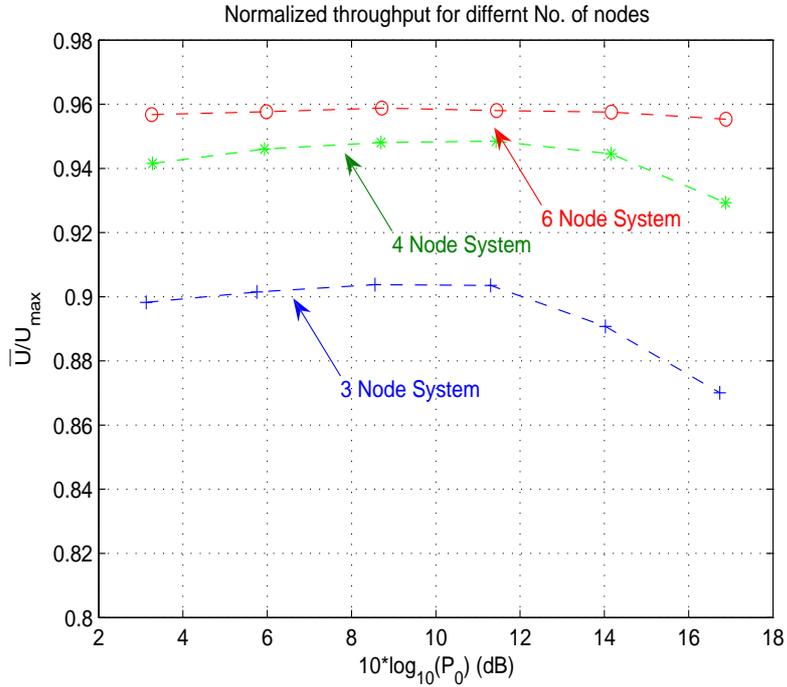}
\caption{ End-to-End normalized throughput of the proposed scheme
(normalized by the strictly optimal performance obtained from brute
force numerical  optimization) versus transmit SNR $P_0$ for N=3,4,6
cognitive relay nodes. The PU activity is given by $\Pr(A_m=0)=0.15$
and the path loss exponent is given by $3$. }\label{fig:Main_node}
\end{figure}

Figure~\ref{fig:Main_node} illustrates the normalized throughput
$\frac{\bar{U}}{U_{max}}$ versus the average transmit SNR ($P_0$)
for various number of cognitive relay nodes where
$\frac{\bar{U}}{U_{max}}$ is obtained from brute-force numerical
optimization of Problem 1. With $N=6$, we have over $95\%$ of the
optimal performance. This illustrates that the proposed scheme is
not only order-optimal but achieves close-to-optimal performance
even in small to moderate number of cognitive relay nodes.

\section{Summary}
In this paper, we have derived  a low complexity hop selection and
dynamic power control policies  to maximize the average end-to-end
throughput of the cognitive multi-hop SU system with dynamic spatial
reuse. By exploiting the time-scale difference between the PU
activity and the CSI dynamics, we decompose the problem into a
\emph{Master problem} and several \emph{Sub Problems}. The solution
obtained is decentralized in the sense that each node determines its
next hop and transmit power based on the local and causal CSI only.
The solution consists of an offline recursion and an online
algorithm with worst case complexity ${\cal O}(M^2)$ and worst case
memory requirement ${\cal O}(M^3)$. Furthermore, the solution is
asymptotically optimal for large number of nodes. Significant
throughput performance has been demonstrated.

\appendices
\section{Proof of Lemma \ref{Lem:flow-balance}}
\label{PLem:flow-balance}
\begin{eqnarray}
 \nonumber\bar{U}_{m} - \bar{U}_{m+1}&=&
\sum_{i=0}^{m-1}\sum_{j=m}^{M}\Pr(i,j) U_{ij} -
\sum_{i=0}^{m}\sum_{j=m+1}^{M}\Pr(i,j) U_{ij}
\\ \nonumber &=& \sum_{i=0}^{m-1}\Pr(i,m) U_{i,m} -
\sum_{j=m+1}^{M}\Pr(m,j)U_{m,j} \;\;\;\;\; \forall m \in \{1 ,...,
M-1\}
\end{eqnarray}

Hence:
\begin{eqnarray}
\nonumber U_m \ge U_{m+1}&\Leftrightarrow&
 \nonumber \sum_{i=0}^{m-1}\Pr(i,m) U_{i,m} -
\sum_{j=m+1}^{M}\Pr(m,j)U_{m,j} \ge 0 \\&\Leftrightarrow &
\sum_{i=0}^{m-1}\Pr(i,m) U_{i,m}\ge  \sum_{j=m+1}^{M}\Pr(m,j)
U_{m,j}
\end{eqnarray}

\section{Proof of Lemma \ref{thm:decompose}}
\label{Pthm:decompose} To prove Problem~\ref{prob:mas} and
Problem~\ref{prob:sub} are equivalent to Problem~\ref{prob:org}, we
first prove the following Lemma:
\begin{Lem}
\label{Lem:exchange} Define:
\begin{eqnarray}
V&=&\max_{\mathbf{X}}\min_{m\in\{1,2...M\}}\left(\sum_{i=1}^{L}A_{mi}f_i(x_i)\right)\label{eqn:v_1}
\\V'&=&\min_{m\in\{1,2...M\}}\max_{\mathbf{X}}\left(\sum_{i=1}^{L}A_{mi}f_i(x_i)\right)\label{eqn:V_2}
\end{eqnarray}
where $\mathbf{X}=\{x_i\in \mathbb{C}_i,i\in\{1,2...L\}\}$ are a set
of independent variables. If $f_i(x_i)$ is finite and $\forall
m\in\{1,2...M\}, i\in\{1,2...L\}$, then:
\begin{eqnarray}
V=V'=\min_{m\in\{1,2...M\}}\left(\sum_{i=1}^{L}A_{mi}f^*_i\right)\label{eqn:exchange_result}
\end{eqnarray}
where $f^*_i=\max_{x_i\in\mathbb{C}_i}f_i(x_i)$.
\end{Lem}
\proof In general, switching of "$\max$" and "$\min$" is not allowed
but there are two specific structures in Lemma~\ref{Lem:exchange}.
that we are exploiting.
\begin{itemize}
\item {\bf Independency Property:} $f_i(x_i), \forall i$ are mutually independent (i.e. they are not coupled by any common
variables), as $\mathbf{X}=\{x_i,i\in\{1,2...L\}\}$ is a set of
independent variables.
\item {\bf Monotone Property:} Since for every $m$ and $i$, $A_{mi}\ge 0$:  $\forall m,i$,
$\sum_{i=1}^{L}A_{mi}f_i(x_i)$ is an non-decreasing function of
$f_i(x_i)$. As a result, $V$ is an non-decreasing function of
$f_i(x_i)$, $\forall i\in\{1,2...L\}$.
\end{itemize}

{\color{deep-blue}
Since $V$ is an non-decreasing function of
$f_i(x_i)$ (\emph{Monotone Property}), $f_i(x_i)\le f^*_i$, $\forall i\in\{1,2...M\}$:}
\begin{eqnarray}
V\le
\min_{m\in\{1,2...M\}}\left(\sum_{i=1}^{L}A_{mi}f^*_i\right)\label{eqn:v_le}
\end{eqnarray}

{\color{deep-blue}
Moreover, denote $x^*_i=\arg\max_{x_i\in\mathbb{C}_i}f_i(x_i)$, from the \emph{Independency Property}, $\{x_i=x^*_i, i\in\{1,2...M\}\}$ is a feasible point for $V$. Hence:}
\begin{eqnarray}
V\ge
\min_{m\in\{1,2...M\}}\left(\sum_{i=1}^{L}A_{mi}f^*_i\right)\label{eqn:v_ge}
\end{eqnarray}

Combining \eqref{eqn:v_le}, \eqref{eqn:v_ge}
$V=\min_{m\in\{1,2...M\}}\left(\sum_{i=1}^{L}A_{mi}f^*_i\right)$.

On the other hand, since $\forall m$,
$\max_{\mathbf{X}}\left(\sum_{i=1}^{L}A_{mi}f_i(x_i)\right)=\sum_{i=1}^{L}A_{mi}f^*_i$:
\begin{eqnarray}V'=\min_{m\in\{1,2...M\}}\left(\sum_{i=1}^{L}A_{mi}f^*_i\right)
\end{eqnarray}
\endproof

In Problem~\ref{prob:org}, for a fixed $\mathcal{P}_{main}$, denote:
\begin{eqnarray}
\nonumber \overline{U}(\mathcal{P}_{main})&=&\max_{{\mathcal
L},{\mathcal P}|\mathcal{P}_{main}}\min_{m\in\{1,..., M\}}
\left(\sum_{i=0}^{m-1}\sum_{j=m}^{M}\Pr(i,j){U}_{ij}\right)
\\&=&\max_{{\mathcal
L},{\mathcal P}|\mathcal{P}_{main}}\min_{m\in\{1,..., M\}}
\left(\sum_{i=0}^{M-1}\sum_{j=1}^{M}A(i,j,m){U}_{ij}\right)
\\\nonumber \mbox{where:}&& A(i,j,m)=\left\{
\begin{array}{l}\Pr(i,j)\;\;\mbox{if: } i<m\le j;
\\0\;\;\;\;\mbox{else} \end{array}\right.
\end{eqnarray}

Note that: a) From (\ref{eqn:av-thp-sub1}), \eqref{eqn:av-subpower},
$U_{ij}$ and $\overline{P}_{ij}$ depends on different set of
variables ${\cal L}_{ij}$ and ${\cal P}_{ij}$. Hence, for a given
${\cal{P}}_{main}=\{\overline{P}_{ij}\}$, constraint
\eqref{eqn:power_2} is decoupled and $\{\mathcal{L}_{ij},
\mathcal{P}_{ij}\}$ become independent variables for different
$\{i,j\}$.

b) From \eqref{eqn:obj_2}, for all $i,j$, $\Pr(i,j)\ge 0$,
$A(i,j,m)\ge0$, $\forall i,j,m$.

Combining a) and b), we can apply Lemma~\ref{Lem:exchange} and
obtain:
\begin{eqnarray} \label{eqn:exchange1}\overline{U}({\mathcal P}_{main})&=&\min_{m\in\{1,...,
M\}}\left( \sum_{i=0}^{m-1}\sum_{j=m}^{M}\Pr(i,j)
U^*_{ij}(\overline{P}_{ij})\right)
\end{eqnarray}
where $U^*_{ij}(\overline{P}_{ij})$ is given by the solution of
Problem~\ref{prob:sub}. Hence, we can rewrite the objective function
as:
\begin{eqnarray}\overline{U}=\max_{{\mathcal P}_{main}}\overline{U}({\mathcal P}_{main})=\max_{{\mathcal P}_{main}}\min_{m\in\{1,...,
M\}}\left( \sum_{i=0}^{m-1}\sum_{j=m}^{M}\Pr(i,j)
U^*_{ij}(\overline{P}_{ij})\right)\end{eqnarray} which is exactly
the objective function in Problem~\ref{prob:mas}. Therefore, the
optimal solution given by Problem~\ref{prob:mas} and
Problem~\ref{prob:sub} shall be the same as Problem~\ref{prob:org}.

{\color{deep-blue}
\section{Proof of Lemma \ref{Lem:lagrange}}
\label{pLem:langrange}
We shall prove that the suboptimal solution $\Omega^{LB}_{ij}$ is asymptotically optimal under the two conditions in Lemma~\ref{Lem:lagrange}.
 We shall first prove the following Lemma:

\fbox{\parbox{15.6cm}{
\begin{Lem}
Suppose: 1) For any $\epsilon>0$, there exists a finite $C>0$ such
that when $|s-t|\geq C$, $G_{st} < \epsilon$;  2) $G_{st}\ge
G_{st'}$, $G_{st} \ge G_{s't}$ when $t'\ge t>s\ge s'$. Then:
\begin{eqnarray}
&&\frac{\sum_{k=1}^{j-i}T(\eta_{ij}(k),\Omega_{ij})}{E^{\Omega_{ij}}\sum_{k=1}^{j-i}T(\eta_{ij}(k),\Omega_{ij})}\rightarrow
1 \mbox{ and}
 \label{eqn:chebi2}
\\&& \frac{\sum_{k=1}^{j-i}P(\eta_{ij}(k),\Omega_{ij})}{E^{\Omega_{ij}}\sum_{k=1}^{j-i}P(\eta_{ij}(k),\Omega_{ij})}\rightarrow
1 \mbox{ in probability when } |j-i|\rightarrow \infty
 \label{eqn:chebi3}
\end{eqnarray}
\end{Lem}}}
\proof  We partition the continuous segment $R_i\sim R_j$ into
$R=\lceil\frac{j-i}{C}\rceil$ clusters:
$\mathbb{V}_r=\{i+rC,i+rC+1,...\min(i+r(C+1)-1,j)\}$,
$r\in\{0,1...R-1\}$. As for any $\epsilon>0$, there exists a finite
$C>0$ such that when $|s-t|\geq C$, $G_{st} < \epsilon$, let
$\epsilon\ll\frac{1}{\overline{P}_{ij}}$, we
have:\footnote{Otherwise,
$T(\eta_{ij}(k),\Omega_{ij})=\frac{1}{\log(1+
G_{s_{ij}(k)l_{ij}(k)}(k)
P_{ij}(k))}\sim\mathcal{O}\frac{1}{\epsilon
\overline{P}_{ij}(k)}\rightarrow \infty$}
\begin{eqnarray}l_{ij}(k)-s_{ij}(k)< C,
\forall k\in\{1,2...j-i\} \label{eqn:hop}\end{eqnarray}

Denote
$T_r=\sum_{l_{ij}(k)\in\mathbb{V}_r}T(\eta_{ij}(k),\Omega_{ij})
=\sum_{s_{ij}(k)\neq j,l_{ij}(k)\in\mathbb{V}_r}\frac{1}{\log(1+
G_{s_{ij}(k)l_{ij}(k)}(k) P_{ij}(k))}$, then:
$\sum_{k=1}^{j-i}T(\eta_{ij}(k),\Omega_{ij})=\sum_{r=0}^{R-1}T_r$.
Moreover, from \eqref{eqn:hop}, we have: $1\le |s_{ij}(k)\neq
j,l_{ij}(k)\in\mathbb{V}_r|\le C$. Moreover, as in practice, the
time duration to transmit one bit should be positive and finite,
there should exist $T_{\min},T_{\max}\in\mathbb{R}^+$ such that
$T_{\min}\le T(\eta_{ij}(k),\Omega_{ij})\le T_{\max}$, $\forall
\eta_{ij}(k)$. Hence we have:
\begin{eqnarray}
T_{\min}\le T_r \le C T_{\max},\forall
r\in\{0,1,...R-1\}\label{eqn:range}\end{eqnarray}

 As we shall proof in Appendix~\ref{pLem:cov}, we have the
following results concerning the covariance between $\{T_r\}$:

\fbox{\parbox{15.6cm}{
\begin{Lem}Given: 1) For any $\epsilon>0$, there exists a finite
$C>0$ such that when $|s-t|\geq C$, $G_{st} < \epsilon$ 2)
$G_{st}\ge G_{st'}$, $G_{st}\ge G_{s't}$  when $t'\ge t>s\ge s'$. We
have:\label{lem:cov} $
\mbox{Cov}(T_r,\sum_{s=0}^{r-1}T_s)\le0\label{eqn:cov} $, $\forall
r\in\{1,2...R-1\}$.~\hfill \IEEEQED
\end{Lem}}}

With Lemma~\ref{lem:cov} and \eqref{eqn:range}, we have:
\begin{eqnarray}\nonumber \mbox{Var}(\frac{\sum_{r=0}^{R-1}T_r}{E^{\Omega_{ij}}\sum_{r=0}^{R-1}T_r})
&=&\frac{\sum_{r=0}^{R-1}\mbox{Var}(T_r)+2\sum_{r=1}^{R-1}\mbox{Cov}(T_r,\sum_{s=0}^{r-1}T_s)}{\left(\sum_{r=0}^{R-1}E^{\Omega_{ij}}T_r\right)^2}
\le\frac{\sum_{r=0}^{R-1}\mbox{Var}(T_r)}{R^2T^2_{\min}}
\\&\le&\frac{RC^2T^2_{\max}}{R^2T^2_{\min}}\rightarrow
0 \;\; \mbox{as: } R=\lceil\frac{j-i}{C}\rceil\rightarrow
\infty\label{eqn:variance}
\end{eqnarray}

Substitute \eqref{eqn:variance} into Chebyshev inequality,
\eqref{eqn:chebi2} is proved; \eqref{eqn:chebi3} can also be proved
through similar process. We shall omit the details due to page
limit.
\endproof

From \eqref{eqn:chebi2}, \eqref{eqn:chebi3}:
\begin{eqnarray}
\nonumber \label{eqn:asyn2}&&E^{\Omega_{ij}}\left[\frac{1}{\sum_{k=1}^{j-i}T(\eta_{ij}(k),\Omega_{ij})}\right]
\rightarrow
\frac{1}{\sum_{k=1}^{j-i}E^{\Omega_{ij}}T(\eta_{ij}(k),\Omega_{ij})}\mbox{ and}
\\ \nonumber \label{eqn:asyn2}&&E^{\Omega_{ij}}\left[\frac{\sum_{k=1}^{j-i}P(\eta_{ij}(k),\Omega_{ij})}{\sum_{k=1}^{j-i}T(\eta_{ij}(k),\Omega_{ij})}\right]
\rightarrow
\frac{\sum_{k=1}^{j-i}E^{\Omega_{ij}}P(\eta_{ij}(k),\Omega_{ij})}{\sum_{k=1}^{j-i}E^{\Omega_{ij}}T(\eta_{ij}(k),\Omega_{ij})}\mbox{ in probability when } j-i\rightarrow
\infty
\end{eqnarray}

Hence, for sufficiently large $j-i$, Problem~\ref{prob:sub} can be
equivalently rewritten as:
\begin{eqnarray}
&&\min_{\Omega_{ij}}\sum_{k=1}^{j-i}E^{\Omega_{ij}}T(\eta_{ij}(k),\Omega_{ij})
\\&\mbox{S.t.:}&\sum_{k=1}^{j-i}E^{\Omega_{ij}}\left(P(\eta_{ij}(k),\Omega_{ij})-\overline{P}_{ij}T(\eta_{ij}(k),\Omega_{ij})\right)\le0
\end{eqnarray}

Observe that the Lagrangian dual function of the above problem is
exactly (\ref{eqn:LB3}). Hence, $U^{LB}(\overline{P}_{ij})
\rightarrow U^*_{ij}(\overline{P}_{ij})$ for sufficiently large
$j-i$.

\section{Proof of Lemma \ref{lem:cov}}
\label{pLem:cov} We shall first prove the following Lemma:

\fbox{\parbox{15.6cm}{
\begin{Lem}\label{Lem:ab} Given three sequences $a_0\le a_1\le ...\le a_N$,
$b_0\ge b_1\ge ...\ge b_N$, $p_n \ge0, n\in\{0,1...N\}$ which
satisfy: $\sum_{n=0}^{N}p_n=1$,
$\sum_{n=0}^{N}p_na_n=\sum_{n=1}^{N}p_nb_n=0$, we have:
$\sum_{n=0}^{N}p_na_nb_n\le0$.~\hfill \IEEEQED
\end{Lem}}}
\proof Denote $N^-_{a}=|\{n:a_n<0\}|$, $N^+_{b}=|\{n:b_n>0\}|$,
where $|\mathbb{A}|$ means the cardinality or set $\mathbb{A}$. If
$N^-_a=N^+_b$, then obviously $\sum_{n=0}^{N}a_nb_n\le0$; Otherwise,
without loss of generality, assume $N^-_a>N^+_b$ and then:
\begin{eqnarray}
\nonumber
\sum_{n=1}^{C}p_na_nb_n=\sum_{n=0}^{N^+_b-1}p_na_nb_n+\sum_{n=N^+_b}^{N^-_a-1}p_na_nb_n+\sum_{n=N^-_a}^{N}p_na_nb_n\le\sum_{n=0}^{N^+_b-1}p_na_nb_n+\sum_{n=N^+_b}^{N^-_a-1}p_na_nb_n
\\\le a_{(N^+_b-1)}\sum_{n=0}^{N^+_b-1}p_nb_n+a_{(N^+_b)}\sum_{n=N^+_b}^{N^-_a-1}p_nb_n \le
(a_{(N^+_b-1)}-a_{(N^+_b)})\sum_{n=0}^{N^+_b-1}p_nb_n\le0
\label{eqn:loser}
\end{eqnarray}
\endproof

Recall the system state transition kernel:
\begin{eqnarray}\Pr(\eta_{ij}(k)|\eta_{ij}(k-1),\Omega_{ij})=\mathbf{1}\left(s_{ij}(k)=l_{ij}(k-1))\right)\Pr(\mathbf{G}_{s_{ij}(k)})
\end{eqnarray}

It can be observed that conditioned on the source node at the frame
$k$: $s_{ij}(k)$, $\{\eta_{ij}(k),\eta_{ij}(k+1),...\}$ are
independent of $\{\eta_{ij}(1),\eta_{ij}(2),...\eta_{ij}(k-1)\}$.
Correspondingly, $\{T(\eta_{ij}(s),\Omega^{ij}),s\in\{k,k+1...\}\}$
are conditionally independent of
$\{T(\eta_{ij}(s),\Omega^{ij}),s\in\{1,2...,k-1\}\}$. Denote
$l_{r\_\min}=\min(l_{ij}(k):l_{ij}(k)\in\mathbb{V}_r)$. Then:
conditional on $l_{r\_\min}$, $T_r$ is independent of
$\{T_s,s\in\{0,1,...r-1\}\}$. Hence:
\begin{eqnarray}
E^{\Omega_{ij}}\left.\left(T_r\cdot\sum_{s=1}^{r-1}T_s\right|l_{r\_\min}=x\right)=E^{\Omega_{ij}}(T_r|l_{r\_\min}=x)E^{\Omega_{ij}}\left.\left(\sum_{s=1}^{r-1}T_s\right|l_{r\_\min}=x\right)\label{eqn:ind}
\end{eqnarray}
where $x\in\{i+rC,i+rC+1,...i+rC+|\mathbb{V}_r|-1\}$. Denote
$k_{r\_\min}=\min(k:l_{ij}(k)\in\mathbb{V}_r)$. Since $G_{st}\ge
G_{st'}$ when $s< t\le t'$, $T(\eta_{ij}(k_{r\_\min})-1)$ is an
non-decreasing function of $l_{r\_\min}$. Correspondingly,
$E^{\Omega_{ij}}\left.\left(\sum_{s=1}^{r-1}T_s\right|l_{r\_\min}\right)$
is a non-decreasing function of $l_{r\_\min}$. Similarly, as
$G_{st}\ge G_{s't}$ when $s'\le s< t$,
$E^{\Omega_{ij}}(T_r|l_{r\_\min})$ is a non-increasing function of
$l_{r\_\min}$. Let
$E^{\Omega_{ij}}\left.\left(\sum_{s=1}^{r-1}T_s\right|l_{r\_\min}=x\right)-E^{\Omega_{ij}}\left(\sum_{s=1}^{r-1}T_s\right)=a_x$,
$E^{\Omega_{ij}}(T_r|l_{r\_\min}=x)-E^{\Omega_{ij}}(T_r)=b_n$,
$\Pr(l_{r\_\min}=x)=p_n$ and substitute to Lemma~\ref{Lem:ab}: $
\sum_{x=i+rC}^{i+rc+|\mathbb{V}_r|-1}\Pr(l_{r\_\min}=x)\left(E^{\Omega_{ij}}(T_r|l_{r\_\min}=x)-E^{\Omega_{ij}}(T_r)\right)
\left(E^{\Omega_{ij}}\left.\left(\sum_{s=1}^{r-1}T_s\right|l_{r\_\min}=x\right)-E^{\Omega_{ij}}\left(\sum_{s=1}^{r-1}T_s\right)\right)\le0
$.  From this result and \eqref{eqn:ind}:
\begin{eqnarray}
\nonumber
&&\mbox{Cov}(T_r,\sum_{s=0}^{r-1}T_s)=E^{\Omega_{ij}}\left(T_r\sum_{s=1}^{r-1}T_s\right)-E^{\Omega_{ij}}\left(T_r\right)E^{\Omega_{ij}}\left(\sum_{s=1}^{r-1}T_s\right)
\\\nonumber &&=\sum_{x=i+rC}^{i+rc+|\mathbb{V}_r|-1}\Pr(l_{r\_\min}=x)\left(E^{\Omega_{ij}}(T_r|l_{r\_\min}=x)-E^{\Omega_{ij}}(T_r)\right)
\\&&\;\;\;\;\cdot\left(E^{\Omega_{ij}}\left.\left(\sum_{s=1}^{r-1}T_s\right|l_{r\_\min}=x\right)-E^{\Omega_{ij}}\left(\sum_{s=1}^{r-1}T_s\right)\right)\le0
\end{eqnarray}
}

\section{Proof of Lemma \ref{Lem:subconcave}}
\label{pLem:subconcave} Due to the {\em Theorem of Lagrangian}
(\cite{FirstOpt}, section 5.2.3), we have
\begin{equation}
\label{eqn:sub_grad} \frac{\partial U^{LB}_{ij}}{\partial
\overline{P}_{ij}}=\lambda^*_{ij}(\overline{P}_{ij})
\end{equation}
where $\lambda^*_{ij}(\overline{P}_{ij})$ is the Lagrange
{\color{deep-green} multiplier} obtained in the subproblem via Algorithm
\ref{Alg:sol_sub}. Hence, Lemma \ref{Lem:subconcave} holds if and
only if $\lambda^*_{ij}(\overline{P}_{ij})$ is a non-increasing
function of $\overline{P}_{ij}$.  Note that in (\ref{eqn:opt_p_sub}), $\forall
k,l_k,G_{s_{ij}(k)l_{ij}(k)}(k)
>0$:  $P_{ij}(k)$ decreases as $\lambda^*_{ij}$ increases. Substitute
this result to (\ref{eqn:power_sub_3}) and it is obvious that
$\lambda^*_{ij}(\overline{P}_{ij})$ decreases as $\overline{P}_{ij}$
increases.

\begin{figure}[ht]
\centering
\includegraphics[scale = 0.8]{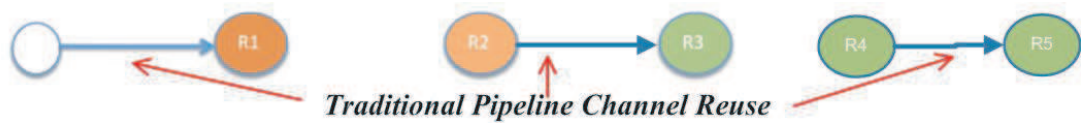}
\caption{Illustration of the traditional "regular pipeline spatial reuse"  relay protocol in a multi-hop network.}\label{fig:relay
scheme_old}
\end{figure}

\begin{figure}[ht]
\centering
\includegraphics[width=4.4in, height = 2.9in ]{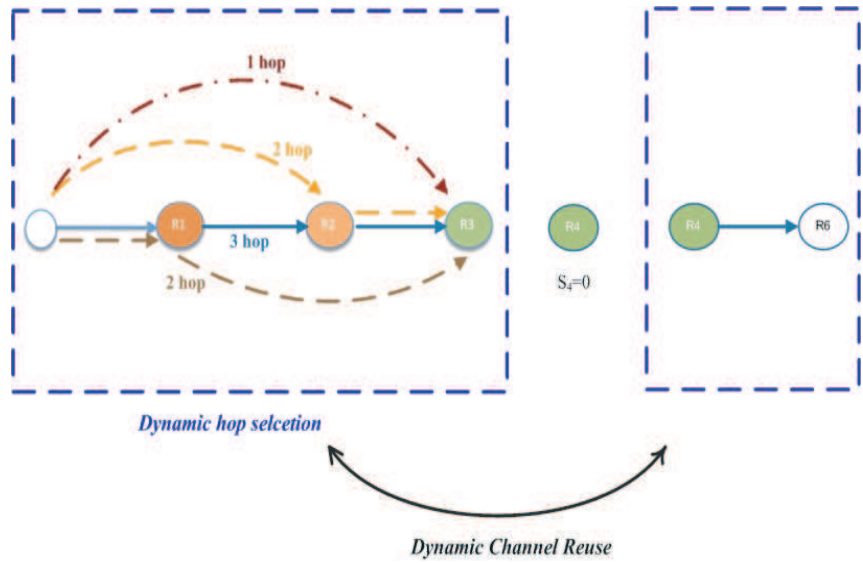}
\caption{Illustration of dynamic spatial reuse when the multi-hop relay chain is partitioned into two {\em continuous segments} by some PU activity realization. We adopt dynamic hop selection within each {\em continuous segment}. }\label{fig:relay scheme}
\end{figure}

\end{document}